\documentclass[nofootinbib, showkeys]{revtex4}

\usepackage{amsmath}
\usepackage{graphics}
\usepackage{graphicx}
\usepackage{dcolumn}
\usepackage{amssymb}
\usepackage{amsthm}
\usepackage{amsfonts}

\usepackage[colorlinks=true,linkcolor=red,citecolor=blue,urlcolor=cyan,
bookmarks=true,bookmarksopen=true,pdfpagemode=None,pdfstartview=FitH]{hyperref}

\newcommand{\be}{\begin{equation}}
\newcommand{\ee}{\end{equation}}
\newcommand{\bea}{\begin{eqnarray}}
\newcommand{\eea}{\end{eqnarray}}

\def\lapp{\mathrel{\rlap{\raise.5ex\hbox{$<$}}
                    {\lower.5ex\hbox{$\sim$}}}}
\def\gapp{\mathrel{\rlap{\raise.5ex\hbox{$>$}}
                    {\lower.5ex\hbox{$\sim$}}}}

\def\M0{m_{0}}

\def\MSTAU1{m_{\tilde \tau_1}}
\def\MXI10{m_{\tilde \chi_1^0}}
\def\MST1{m_{\tilde t_1}}

\begin{document}


\title {Comments on the reach of INO experiment: JHEP {\bf 1304}, 009 (2013) and   JHEP {\bf 1305}, 058 (2013)
}
\author
{\sf Abhijit Samanta\footnote{E-mail address: abhijit.samanta@gmail.com}
}
\affiliation
{{\em Department of Physics, Heritage Institute of Technology, Kolkata 700 107, India 
}}
\begin{abstract}
{
In JHEP {\bf 1304}, 009 (2013) and JHEP {\bf 1305}, 058 (2013) 
the {\it reach of INO experiment} for determination of  neutrino mass hierarchy and the sensitivity to both $\Delta m_{32}^2$ and $\theta_{23}$ have been reported, which are significantly underestimated  and drastically different from earlier studies \cite{as-jhep,as-hier-prd,as-sen} and strongly dependent on the flux uncertainties. Here, we clarified that the effect on oscillation probability due to change of oscillation parameters are not  considered appropriately due to improper binning of data, reconstruction of muon energy and angular resolutions from  events in together with fully contained events and partially contained events,  improper incorporation of resolutions, 
and rejection of high energy events $E\gapp $ 10 GeV.   
} 
\end{abstract}

\maketitle

\section{Introduction} 
There are several studies \cite{refhier,refsen} in the last few years to estimate the potential for measurements of neutrino oscillation parameters at  future atmospheric neutrino experiments, particularly, using a large magnetized iron calorimeter detector proposed at India-based Neutrino Observatory \cite{ino}.   The results are significantly different from one to another and there is almost no clarification why they differ significantly.  Recently, in \cite{hier} and \cite{sen}, {\it the reach of  this experiment for the measurement of the atmospheric neutrino parameters has been reported.} The  results are significantly underestimated and drastically different from earlier analysis \cite{as-hier-prd,as-jhep}. For an example, for input inverted hierarchy (IH), $\theta_{23}=45^\circ$ and $\theta_{13}=8.23^\circ$ ($\sin^22\theta_{13}=0.08$),  the sensitivity to discrimination of mass hierarchy ($\Delta \chi^2$) for 20 years of INO data is  8.5 without marginalization and  flux uncertainties \cite{hier}  (see figure 4 in \cite{hier}) and 13.5 with marginalization in \cite{as-hier-prd} (see figure 5 in \cite{as-hier-prd} keeping in mind that the marginalization  range is wider than present $3\sigma$ range). The goal of this paper is to point out the factors in the analysis techniques, which lead to the large differences. 

\begin{figure*}[htb]
\includegraphics[width=8.5cm,angle=0]{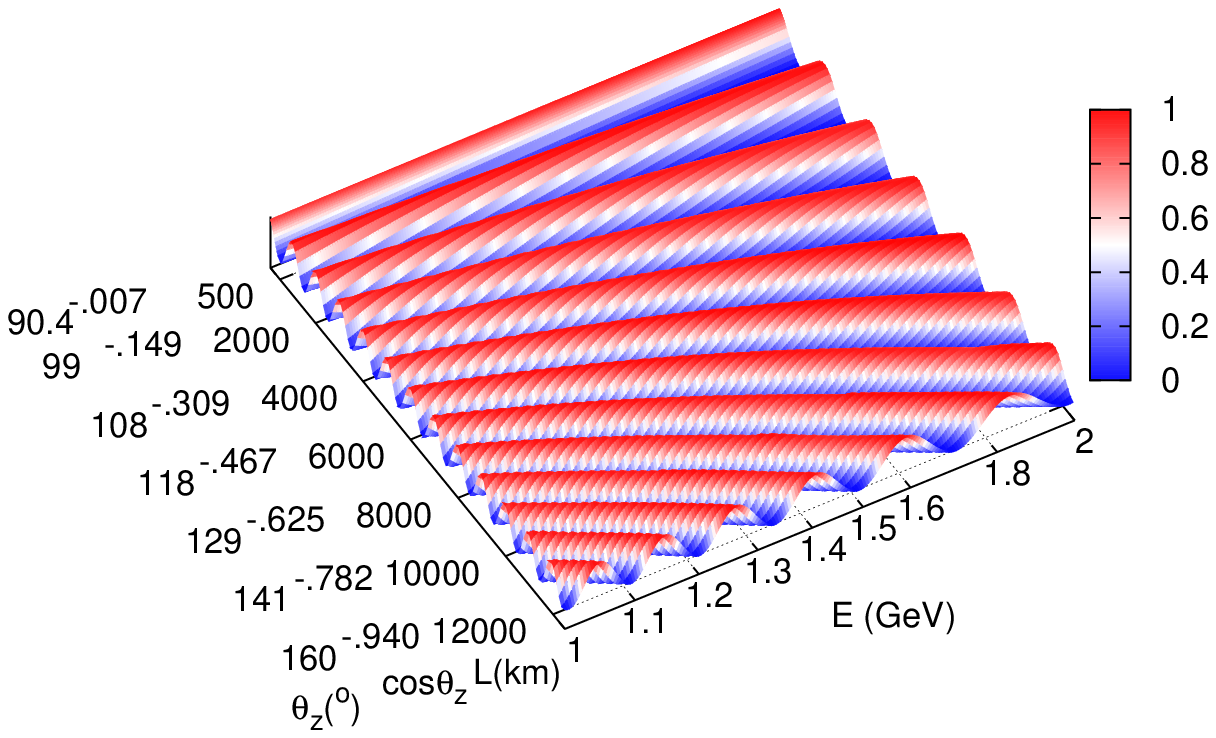}
\includegraphics[width=8.5cm,angle=0]{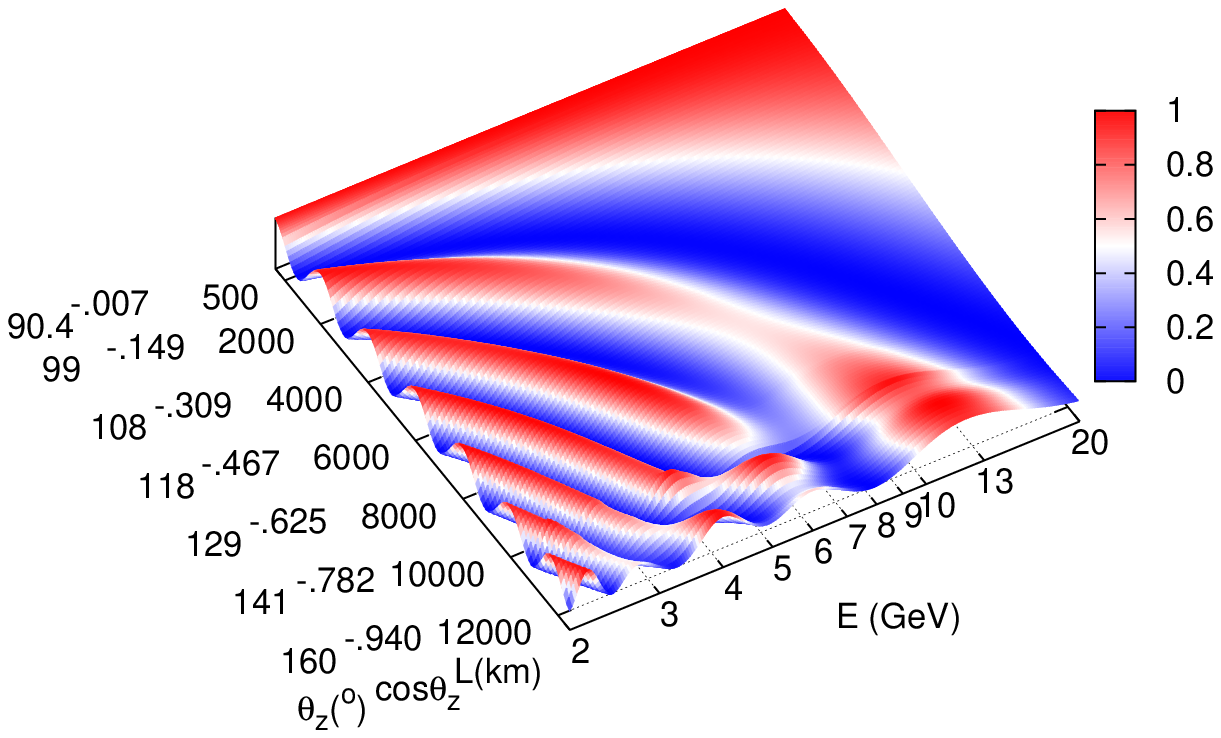}
\caption{\sf \small
The oscillogram for neutrino survival probability $P(\nu_\mu \to \nu_\mu)$ with inverted hierarchy (IH) for $E_\nu = 1-2$ GeV (right), and $2-20$ GeV (left), respectively.
We set $|\Delta m_{32}^2| = 2.5\times 10^{-3}$ eV$^2$, 
$\theta_{23}=45^\circ$, $\theta_{13}=10^\circ$ and $\delta_{CP}=180^\circ$.
}
\label{f:pmumu}
\end{figure*}
\section{Controlling factors}
\subsection{Binning of data}
The result of a statistical analysis of the experimental data 
 depends strongly on the method of binning of the data \cite{as-com, as-sen}. In this experiment the survival probability $P_{\mu\mu}$  as a function of baseline ($L$) and energy $(E)$ will be measured in terms of number of events for $\nu_\mu$ and $\bar\nu_\mu$ separately. The variation of $P_{\mu\mu}$ in $L-E$ plane is shown in figure \ref{f:pmumu}  for a given set of oscillation parameters:
$\Delta m_{32}^2, \Delta m_{21}^2, \theta_{23}, \theta_{13}, \theta_{12}$, and $\delta_{CP}$. The binning of the data should be such that it should produce maximum  sensitivity to $P_{\mu\mu}$ with the changes of oscillation parameters. 
To find the sensitivity of one parameter which even may be sensitive only in a small part of $L-E$ space,  one needs to marginalize the $\chi^2$ over the whole allowed ranges of all other oscillation parameters as they are uncertain over the allowed ranges; and the whole region in $L-E$ plane contributes to its sensitivity due to dependence of other oscillation parameters on whole $L-E$ region. For an example, 
the   mass hierarchy is sensitive to some  small regions of $L-E$ plane (see figure \ref{f:pmumu} and also \cite{as-hier-prd, as-hier-plb}); but,  $|\Delta m_{32}^2|$ and $\theta_{23}$,  are sensitive over whole $L-E$ plane.


The distance between two consecutive oscillation peaks $D_E(L)$ ($D_L(E))$ along $L(E)$-axis for a given fixed value of $E(L)$ increases with increase in $E$ (see figure \ref{f:pmumu}), which again changes  with the change of oscillation parameters. If the bin size is equal or bigger than this distance, the oscillation effect is averaged out and the sensitivity to the oscillation parameters falls significantly.  It requires varying (gradually decreasing) bin size with decrease in $E$. Even if the resolution is worse (bin size $\lapp$ resolution width), the result would improve with decrease in bin size. The improvement may be very small after certain bin size, but would not create any problem unless the number of events in a bin is less than the number of minimum required events  ($n_{ev}^{min}$). This is required  to have $\chi^2$ per degrees of freedom $\approx 1$. 

There are six oscillation peaks in $E$ range 1 -- 2 GeV for  $L=12000$ km; but, if one considers only 1 or 2 bin    (as done in \cite{hier, sen}),  the oscillation effect is then fully averaged out. In \cite{hier,sen} the bin size for $E$ is 1 GeV,  the oscillation effect is averaged out for $E \lapp 4$ GeV, and the crisis of number of events in a bin begins to maintain number of events  $\gapp n_{ev}^{min}$ for $E\gapp 4$, which becomes serious when $E\gapp$ 10 GeV. 

On the other hand, if the binning is done with equal bin size in $\log E$, the problem of averaging out of oscillation probability is reduced for $E\lapp$ 4 GeV and the number of events per bin does not reduce drastically for $E\gapp 4$ GeV  (flux $\sim E^{-\gamma}, \gamma \sim 3$). One can now consider all high energy events ($E\gapp 10$ GeV)  in the analysis (as the bin size is large for high $E$), which would increase the sensitivity drastically in spite of less number of events, but due to high angular resolutions (see resolutions due to kinematics of scattering in \cite{as-sen}).   The sensitivity ($\chi^2$) to oscillation parameters  will be maximized when the binning is done with equal bin size in $\log E$ and and the sensitivity will increase significantly  even with only 10 bins (10 bins in $E$ are considered in\cite{hier, sen}). But, decreasing bin size  with equal binning in $E$ could not improve the sensitivity significantly as the bin size can not be made less than $D_L(E)$ for $E\lapp 2$ GeV maintaining number of events $\gapp n_{ev}^{min}$ for $E\gapp 4$ GeV.      

Again, from figure \ref{f:pmumu}, it is clear that matter effect is not negligible for $E\gapp $10 GeV. Here, the distance between two peaks in $E\approx$ 10GeV for $L=12000$ km and it increases for lower values of $L$. Obviously, one can easily consider bin size in $E$ much much higher than 1 GeV, and then number of events in a bin will $\gapp n_{ev}^{min}$ and sensitivity will increase significantly. 


\subsection{Resolutions}
The muon  energy and angle resolutions at magnetized iron calorimeter (ICAL) are very good over almost whole $L-E$ regions except a small region at the near horizon. However, these resolutions are very negligible compared to the  resolutions due to kinematics of the neutrino scattering processes for whole region of $L-E$ plane and even at near horizon (see energy angle correlated resolutions in the plane consists of $(E_\nu-E_\mu)/E_\nu$ and $(\theta_\nu^{z} - \theta_\mu^z)$ in \cite{as-sen} and compare it with the resolutions of muons for ICAL in \cite{hier,sen}).

 The energy resolution can not fully average out the oscillation pattern as $D_L(E)$ is larger than the resolution width in ($E_\nu-E_\mu$) for whole range of $E$ above 1 GeV, even with considering muon resolution of ICAL detector except the region at near horizon.
For $E_\nu \sim $ 1 GeV, $D_L(E)$ is very small; but,  $(E_\nu-E_\mu)/E_\nu$ is also very small ($\lapp 10-20\%$) as events are mostly from quasi-elastic process (see figure 4 in \cite{as-sen}). If the bin size  in $E$  is less than $D_L(E)$, it  will contribute significantly to the sensitivity to the oscillation parameters. 

On the other hand, though the angular resolution width in $(\theta_\nu^{z} - \theta_\mu^z)$ is large at $E_\nu\sim$ 1 GeV, but it decreases very rapidly with increase in $E_\nu$  and resolutions in $(\theta_\nu^{z} - \theta_\mu^z)$ becomes smaller than the distance between two peaks in $\theta_\nu^z$  for $E_\nu \gapp 1.5$ GeV. Again, for events where muons are going near vertically, the change of $L$  ($\cos\theta^z$) with $\theta^z$ is relatively small around the vertical axis than the change in the near horizon (see figure \ref{f:pmumu}), but the resolution $\theta_\nu-\theta_\mu$ (due to kinematics) remains same. 

Moreover, for event going vertically the distribution of the event  on the both side of the vertical axis due to smearing of resolution will fall twice in a $L$ bin (equal angles around the vertical axis produce same $L$). In spite of wide angular  resolutions, this will enhance the sensitivity compared to the events far away from this zone.        

There is 7\% increase of events after incorporation of resolutions in \cite{sen, hier} (see table 3 in \cite{sen}), which may shift the best-fit values significantly from their true values. In \cite{hier,sen} the resolutions are incorporated bin wise, the number of events of a bin is smeared using Gaussian resolution function considering the central value of the bin. 
Again. the number of $\cos\theta^z$ bins for smearing resolutions as well as $\chi^2$ analysis are same; which reduces 
the effect of resolution from the actual one.


The angular resolution function has been constructed in $\cos\theta^z$ in \cite{hier,sen}. It would not be a fully Gaussian for nearly vertical events, it will ends at $\cos\theta^z=-1$.  But, in \cite{sen,hier} the smearing has been done using a  Gaussian resolution in $\cos\theta_{\rm z}$  considering standard deviation $\sigma_{\cos\theta_z}$ obtained from GEANT-based simulation ( see eq. 7 of \cite {sen}). The smearing of the event by integrating over $-\infty$ to $-1$ and adding the contribution to the last bin will overestimate the number of events in that bin and underestimate  in all other bins. This will distort the event distribution and misplace the best-fits of oscillation parameters from their true values due to large tail of Gaussian functions as the resolution width is not always much smaller than the bin size.

The muon looses energy due to mainly  ionization and atomic excitation for $E\sim$ a few GeV. So, in case of fully contained (FC) events the path length traversed in a medium is  proportional to its energy. On the other hand, in magnetic field the muon track  bends with increasing curvature along  its track as its energy decreases gradually. In case of FC events one can measure the  energy from effective path length (density times path length) as well as from curvature.  The measurement of curvature depends strongly on precision  of the determination of hit positions  in the active elements of the detector.  This dependence is relatively less in case of measurement of track length. So, it is highly expected that {\it the energy resolution from track length will be much better than from curvature.}   Moreover, as muon energy increases the curvature decreases and resolution width will increase with energy. But,  ($\sigma(E)/E$) is then expected not to  increase with energy (at least for $E \lapp $ a few tens of GeV) for FC events if it measured from effective path length (as the relative error in measurement of track length decreases with its increase in magnitude).

On the other hand, the muon energy for partially contained (PC) events can only be measured from curvature. If the track is not long and/or energy is high (bending is small), the measurement of energy will be poor in compared to FC events.  The muon energy resolutions for PC events would obviously be wider significantly in comparison to the FC events.   

It is not mentioned in the papers \cite{hier,sen} that the fully contained (FC) and partially contained (PC) events are treated separately in construction of resolution functions. As muon energy increases the number of PC events begins to increase and becomes larger than FC events. This may be  one of the main reason why muon energy resolution is significantly worsened for $E\gapp$ 5 GeV in energy resolution plots in \cite{hier, sen}. There would not appear significant difference in muon angular resolutions between FC and PC events.  One can show that with only FC events one can get better sensitivity
than the combined one  with FC and PC
events. The result will improve with PC events only when they are
treated separately. As I find, this fact (which belongs to physics
analysis, not to the construction of resolution functions)  has not
been discussed in $\chi^2$ analysis or in any part of the papers \cite{hier,sen}. 

The same procedure (using smearing of resolutions) is used to generate both experimental data set as well as theoretical data set. If one {\it demands the results as a reach of an experiment or from detector simulation,} then the experimental data set should be obtained from directly reconstructing events. 
In analysis of actual experimental data,  sensitivity as well as the best-fits will be significantly different due to the the above method of incorporation of resolutions.  In my opinion the simplest and best way to do the analysis is without using any resolution function, but directly reconstructing all events of large number of ICAL events (say, 10000 years of data)  for one time and then incorporating oscillation and  reducing the number of events according to the exposure time.   

\subsection{Detector efficiency}  
The muon detection efficiency is $\gapp 90\%$  over almost whole region except at the near horizon where the efficiency is much less and 
 up going and down going events are mixed up due to wide resolutions  and the contribution to the $\chi^2$ is expected to be significantly small compared to the region away from horizon.  This has been discussed in detail with results using different horizontal cuts in sec. VII.D in \cite{as-sen}. The results can not be significantly worsened  due to much less efficiency at near horizon. 

\subsection{Flux and cross section uncertainties}
The effect of flux uncertainties on the sensitivity to oscillation parameters is marginal for $E_\nu > 1$ GeV. For $E_\nu > 1$ GeV, there is mainly overall flux normalization uncertainty and  the effect would be minimized by the pull method of $\chi^2$ if the bin size in both $L$ and $E$ are much less than $D_E(L)$ and $D_L(E)$, respectively. In section 5.2 of \cite{as-jhep} we have shown that the effect of it on the sensitivities to $\Delta m_{32}^2$, $\theta_{23}$, and mass hierarchy are very marginal in contrary with \cite{hier, sen}.

\section{Zones sensitive to each oscillation parameters}
\begin{enumerate}
\item 
{\it $\Delta m_{32}^2$:} The sensitivity to $\Delta m_{32}^2$  comes from the measuring of $D_E(L)$ and $D_L(E)$. It requires bin size in $L$ ($E$) much smaller than $D_E(L)$ ($D_L(E)$).

\item {\it $\theta_{23}$:}  The sensitivity to deviation from maximal mixing  ($|45^\circ - \theta_{23}|$)  comes from whole  region of $L-E$ and the sensitivity to octant ($45^\circ - \theta_{23}$) comes from depleted region as shown in figure \ref{f:pmumu} due to the matter effect (discussed in detail in \cite{as-hier-prd, as-hier-plb}).

\item {\it $\delta_{CP}$:} The sensitivity to $\delta_{CP}$ arises from the events with $E_\nu \lapp 2 GeV$. However, the 
contribution to $\chi^2$ from events $E_\nu \lapp 0.6$ GeV is negligible due to i) tilt uncertainty in flux for $E_\nu < 1$ GeV and ii) drastic increase of scattering angle $\theta_\nu-\theta_\mu$ with decrease in $E$ \cite{as-cp}.   

\item {\it $\theta_{13}$ and mass hierarchy:} The coupling between solar and atmospheric neutrino oscillation occurs through $\theta_{13}$,  then the sensitivity is expected to come from i) $E_\nu \lapp 1$GeV and also from ii) the depleted regions due to matter effect. Due to the issues discussed in above, the sensitivity from events with $E_\nu \lapp 1$GeV is negligible (see \cite{as-sen}).

\end{enumerate}

\section{Conclusion}
The reach of the INO experiment for measurement of oscillation parameters reported in \cite{hier, sen}  are significantly underestimated and drastically different  from previous studies \cite{as-jhep, as-sen,as-hier-prd} due to improper binning of the data, improper incorporation of resolutions,  rejection of high energy events $\gapp $ 10 GeV. 
The effects on oscillation probability  due to the changes of  oscillation parameters are not reflected fully  in the $\chi^2$ analysis.

\end{document}